\documentclass[prl,twocolumn,aps,amssymb,footinbib,showpacs]{revtex4}
\usepackage{amssymb}

\usepackage{graphicx}
\usepackage{amsmath}
\usepackage{times}
\usepackage{color}
\usepackage{subfigure}
\usepackage{setspace}

\begin{document}

\title{Preparation and detection of magnetic quantum phases in optical superlattices}

\author{ A. M. Rey$^{1}$,  V. Gritsev$^{2}$, I. Bloch $^{3}$, E. Demler$^{1,2}$ and M.D.
Lukin$^{1,2}$}  \affiliation{$^{1}$ Institute for Theoretical
Atomic, Molecular and Optical Physics, Harvard-Smithsonian Center of
Astrophysics, Cambridge, MA, 02138.} \affiliation{$^{2}$ Physics
Department, Harvard University, Cambridge, Massachusetts 02138, USA}
\affiliation{$^{3}$  Johannes Gutenberg-Universit\"{a}t, Institut
f\"{u}r Physik, Staudingerweg 7,55099 Mainz, Germany}

\begin{abstract}
We describe a novel approach   to prepare, detect
and characterize magnetic  quantum phases in ultra-cold spinor atoms
loaded in optical superlattices. Our technique makes use of
singlet-triplet  spin manipulations in an array of isolated  double well
potentials in analogy to recently demonstrated quantum control in
semiconductor quantum dots. We  also discuss the many-body singlet-triplet  spin  dynamics
arising from coherent coupling between nearest neighbor double wells
 and  derive an effective
description for such system. We  use it to study the generation of
complex magnetic  states by adiabatic and non-equilibrium dynamics.
\end{abstract}

\pacs{05.50.+q  03.67.Mn  05.30.Fk  05.30.Jp}
 \maketitle

Recent advances in the manipulations of ultra-cold atoms in optical
lattices have opened new possibilities for exploring  complex
many-body systems \cite{Greiner}. A particular topic of continuous
interest is the study of quantum  magnetism    in
 spin
 systems \cite{Auerbch,Stenger,Sadler}. By loading   spinor atoms in optical lattices it is now possible to "simulate" exotic spin
   models  in  controlled
environments and to explore novel spin orders and phases.

In this Letter we describe a new approach for preparation  and probing of many-body magnetic quantum states that
 makes use of coherent manipulation of singlet-triplet  pairs  of ultra-cold atoms  loaded in deep period-two
  optical superlattices. Our approach makes use of  a spin dependent energy offset between the double-well minima to
  completely control and measure the spin state of two-atom pairs, in a
way analogous to the recently demonstrated manipulations of coupled
electrons in semiconductor double-dots \cite{Petta}. As an example,
we show how this technique allows one to detect and analyze
anti-ferromagnetic spin states in optical lattices.  We further study
the many-body dynamics that
 emerge when tunneling between nearest neighbor double wells is allowed.
%We derive   the effective Hamiltonian that describes  singlet and triplet dynamics
As two specific examples,  we show  how  a set of  singlet atomic
states  can be evolved into singlet-triplet cluster-type states and
into a maximally
 entangled superposition  of two anti-ferromagnetic states. Finally, we
 discuss the use of our projection technique to probe
 the density of spin defects (kinks)  in magnetic  states
 prepared  via equilibrium and non-equilibrium dynamics.

The key idea of this work is illustrated by considering
 a pair of ultra-cold  atoms  with two relevant internal states,
 which we  identify with spin up and down $\sigma=\uparrow,\downarrow$
in an isolated double well (DW) potential as  shown in Fig.1. By
dynamically changing the optical lattice parameters, it is possible
to completely control this system and measure it in an arbitrary
two-spin basis. For concreteness, we first focus on the fermionic
case. The physics of this system is governed by three sets of energy
scales: i) the  on-site interaction energy
$U=U_{\uparrow\downarrow}$ between the atoms,
%which is proportional to the s-wave scattering lengths (for fermions
%$ U_{\uparrow\downarrow} \gg U_{\uparrow\uparrow},U_{\downarrow\downarrow}$),
ii) the  tunneling energy of the $\sigma$ species:   $J_\sigma$,
 and iii) the energy difference between the
 two DW minima, $2\Delta_\sigma$ for each of the two species.
The $\sigma$ index in $J$ and $\Delta$ is due to the fact that the
lattice that  the $\uparrow$ and $\downarrow$  atoms feel can be
engineered to be different  by choosing laser beams of
appropriate  polarizations, frequencies, phases and intensities.  In
the following we assume that the atoms are strongly interacting, $U
\gg J_\sigma$, and that effective vibrational energy of each well,
$\hbar \omega_0$, is the  largest energy scale in the system $\hbar
\omega_0\gg U,\Delta_\sigma,J_\sigma$, i.e deep wells.

Singlet $|s\rangle$ and triplet $|t\rangle$ states  form the natural basis for the two-atom system.
   The relative energies of these states
can be manipulated by controlling the energy bias $\Delta_\sigma$
between the two wells. In the unbiased case ($U\gg  2\Delta_\sigma$)
 only states with one atom per site $(1,1)$  are
populated,
 as the large atomic repulsion energetically suppresses double occupancy (here, labels $(m,n)$ indicate
  the integer number of atoms in the left and right
sites of the DW).  For weak tunneling and spin independent lattices
($J_\uparrow = J_\downarrow =J$, $\Delta_\uparrow =
\Delta_\downarrow=\Delta$) the states   $(1,1)|s\rangle$ and
$(1,1)|t \rangle$ are nearly degenerated. The small energy splitting
between  them is $\sim 4J^2/U$, with the singlet  being the low
energy state (Fig. 1a). As  $\Delta$ is increased  the
relative energy of doubly occupied states $(0,2)$  decreases.
Therefore, states $(1,1)|s\rangle$ and $(0,2)|s\rangle$ will
hybridize. When $2 \Delta \gtrsim U$ the atomic repulsion  is
overwhelmed and consequently the $(0,2)|s\rangle$ becomes the ground
state.
%In the intermediate regime, $2 \Delta \approx U$,
At the same time, Pauli exclusion  results in a large energy
splitting $\hbar \omega_0$ between doubly occupied singlet and
triplet states as the latter   must have  an antisymmetric orbital
wave function. Hence, $(1,1)|t\rangle$ does not hybridize with its
doubly occupied counterpart, and its relative energy becomes large
as compared to the singlet state. Thus the energy difference between
singlet and triplet states can be controlled using $\Delta$.

Further control  is provided by   changing  $J_{\sigma}$ and
$\Delta_\sigma$ in  spin dependent lattices (see Fig.1b).
Specifically,  let us now consider the regime $2\Delta_\sigma\ll U$
in which only  $(1,1)$ subspace  is  populated.  Within this manifold
we define \cite{Sachdev}

\begin{eqnarray}
|s\rangle&=&\hat{s}^\dagger|0\rangle\equiv\frac{1}{\sqrt{2}}(
|\uparrow\downarrow\rangle-|\downarrow\uparrow
\rangle),\\
|t_z\rangle&=&\hat{t}_{z}^\dagger|0\rangle\equiv\frac{1}{\sqrt{2}}(|\uparrow\downarrow\rangle+
|\downarrow\uparrow\rangle),\end{eqnarray}
\begin{eqnarray}
|t_x\rangle&=&\hat{t}_{x}^\dagger|0\rangle\equiv
\frac{-1}{\sqrt{2}}(|\uparrow\uparrow\rangle-|\downarrow
\downarrow\rangle),\\
|t_y\rangle&=&\hat{t}_{y}^\dagger|0\rangle\equiv \frac{i}{\sqrt{2}}
(|\uparrow\uparrow\rangle+ |\downarrow\downarrow\rangle)
\end{eqnarray} Here $\hat{t}^\dagger_{\alpha}$ and $\hat{s}$ are
operators that create triplet and singlet states from the vacuum
$|0\rangle$ (state with no atoms). They satisfy bosonic commutation
relations and the constrain
$(\sum_{\alpha=x,y,z}\hat{t}^\dagger_{\alpha} \hat{t}_{\alpha})+
\hat{s}^\dagger \hat{s}=1$, due to the physical restriction that the
state in a double well is either a singlet or a triplet. In the rest of the letter we will omit the label $(1,1)$ for the
singly occupied states.

 When $\Delta_\sigma$ depends on spin, i.e
$\Upsilon\equiv \Delta_\uparrow-\Delta_\downarrow\neq0$,
 the $|t_z\rangle$
component mixes with $|s\rangle$(see Fig.1c). Note that on the other hand  $|t_{x,y}\rangle$
remain decoupled from $|t_z\rangle$ and $|s\rangle$ .  As a result the
states $|s\rangle$ and $|t_z\rangle$ form an effective two-level
system whose dynamics is driven by the Hamiltonian:
\begin{equation}
\hat{H}_{1}^{J}=- \zeta (\hat{s}^\dagger\hat{s}-
\hat{t}_{z}^\dagger\hat{t}_{z})
-\Upsilon \tilde{S}^z + {\rm const},
 \label{efJ}\end{equation} Here $\zeta\equiv 2 J_{\uparrow}
J_{\downarrow}/\tilde{U}$, is the exchange coupling  energy
(with $\tilde{U}\equiv\frac{U^2-(\Delta_\uparrow+\Delta_\downarrow)^2}{U}$)
and $\tilde{S}^z =\hat{s}^\dagger \hat{t}_z +
\hat{t}^\dagger_z \hat{s}$.
%The upper and lower signs are for fermions and bosons respectively .
If $\Upsilon =0$,  exchange dominates and $|s\rangle$  and
$|t_z\rangle  $ becomes the ground and first excited states
respectively. However if $ \Upsilon \gg \zeta$, exchange can be
neglected and the ground state  becomes either
$|\uparrow\downarrow\rangle$ or $|\downarrow\uparrow\rangle$
 depending on the sign of $\Upsilon$.

\begin{figure}[htbp]
\centering
\includegraphics[width=3.5in]{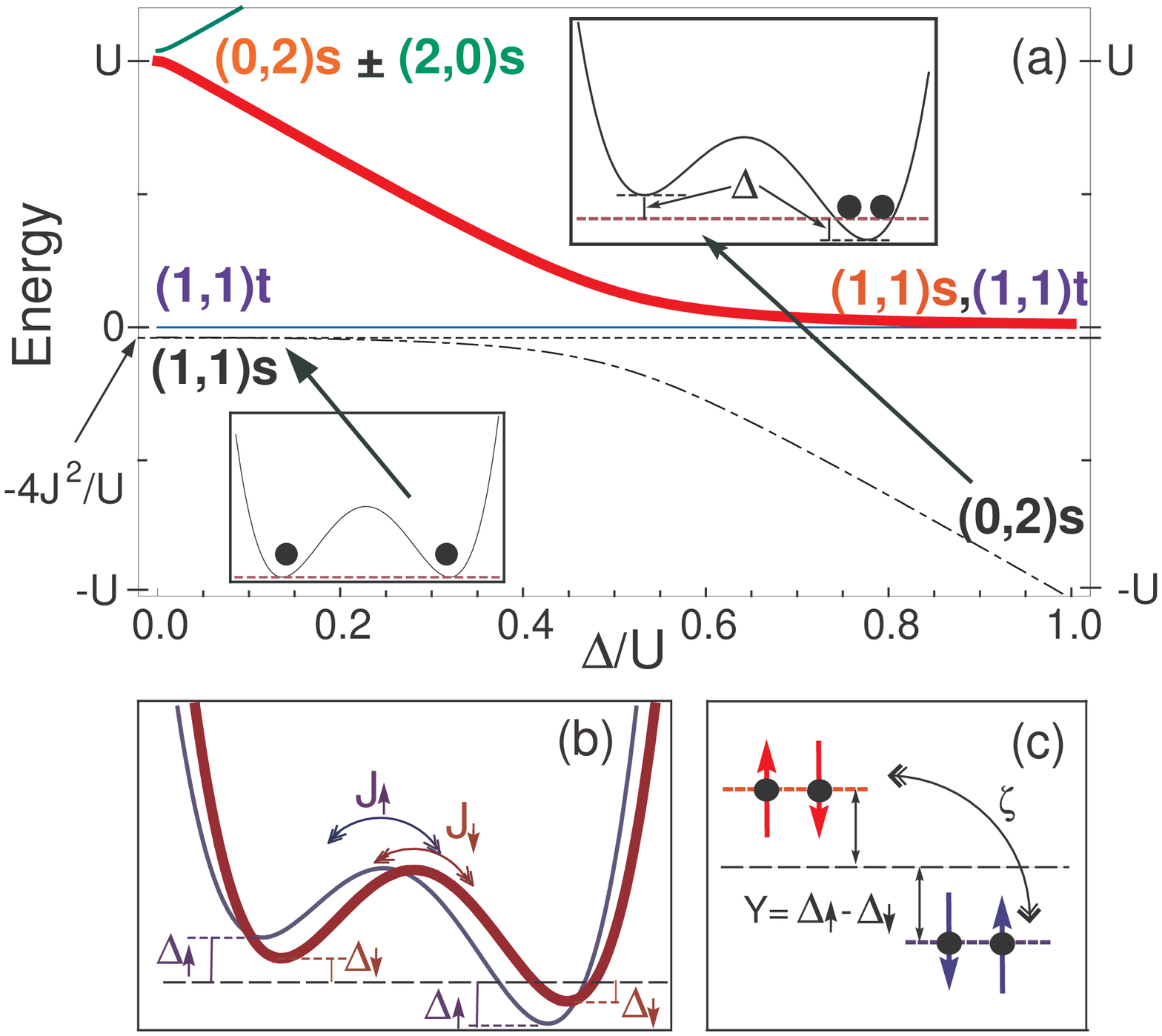}\\
\caption{(color online)
a) Energy levels of fermionic atoms in a spin independent double well as $\Delta/U$ is varied:
While in the regime  $ 2 \Delta \ll U$,  $(1,1)|s\rangle$  is the lowest energy state,  when  $ 2 \Delta \gtrsim U$,
$(0,2)|s\rangle$ becomes the state with lowest energy. b) In spin dependent potentials the two species feel
different lattice parameters c)
Restricted to the $(1,1)$ subspace $\Upsilon$ acts as an effective magnetic field gradient and couples
the  $|s\rangle$ and $|t_z\rangle$
states .}
\end{figure}

These considerations indicate  that it is possible to perform
arbitrary coherent manipulations  and  robust measurement of atom
pair spin states.  The former can be accomplished by combining
time-dependant control over $\zeta, \Upsilon$
to obtain effective rotations on the spin-$1/2$ Bloch sphere within
$|s\rangle-|t_z\rangle $  state. In the parameter regime of interest, $\zeta, \Upsilon$,
can be  varied independently in experiments. In addition, by applying
pulsed (uniform) magnetic fields it is possible to rotate the basis,
thereby changing the relative population of
 the $|t_{x,y,z} \rangle$ states. Atom
pair spin states can be probed  by adiabatically increasing $\Delta$
until it becomes larger than $U/2$, in which case atoms in the
$|s\rangle$ will adiabatically follow  to $(0,2)|s\rangle$ while the
atoms in $|t_\alpha\rangle$ will remain in (1,1) state (Fig. 1a). A
subsequent measurement of the number of doubly occupied wells will
reveal  the  number of singlets in the initial state. Such a
measurement can be achieved by efficiently converting the doubly
occupied wells into molecules via photoassociation  or using other
techniques such as microwave spectroscopy and spin changing
collisions \cite{Bloch}. Alternatively, one can continue
adiabatically  tilting the DW until it merges to one well. In such a
way  the $|s\rangle$ will be projected to the $(0,2)|s\rangle$, while
the triplets will map to  $(0,2)|t_\alpha\rangle$. As
$(0,2)|t_\alpha\rangle$ has one of the atoms in the first
vibrational state of the well, by measuring the population in
excited bands one can detect the number of initial
$|t_\alpha\rangle$ states. Hence the {\it spin-triplet blockade}
\cite{Petta} allows to effectively control and measure atom pairs.

 Detection and diagnostics of many-body spin phases such as antiferromagnetic (AF) states is an example of
 direct application of the singlet-triplet manipulation and measurement technique.
 The procedure to  measure the AF state population is the following; after inhibiting tunneling between the various DWs,
 one can abruptly  increase $\Upsilon$, such that the initial state is projected into
  the new eigenstates $|\uparrow\downarrow\rangle$ and $|\downarrow\uparrow\rangle$ at time $\tau=\tau_0$.
   For  $\tau>\tau_0$ $\Upsilon$ can then  be adiabatically decreased to zero, in which case the
$|\uparrow\downarrow\rangle$ pairs will be adiabatically converted into
$|s\rangle$  and $| \downarrow\uparrow \rangle$ pairs to
 $|t_z\rangle$. Finally, the  singlet population can be measured using the spin blockade. As a result,
   a measure of  the doubly occupied sites (or excited bands population) will detect the number of
    $|\uparrow\downarrow\rangle$ pairs and thus probe   antiferromagnetic
    states of the type $|\uparrow\downarrow\uparrow\downarrow ...\rangle$.

These ideas can be directly
    generalized to perform measurements of the more complex magnetic states that can be represented as
    products of two atom pairs. For example, a pulse of  RF magnetic field can be used to
    orient all spins, thus providing the ability to
    detect $|AF\rangle$ states aligned along an arbitrary direction.
Moreover, one can determine    the relative phase between  singlet
and triplet pairs in $|AF\rangle$ states of the form
$\prod|s\rangle+ e^{i\phi} |t_z\rangle$ by performing Ramsey-type
spectroscopy. After letting the system  evolve freely  (with
$\Upsilon=0$) so that  the $|s\rangle$ and $|t_z\rangle$ components
accumulate  an  additional relative phase due to exchange,  a
read-out pulse (controlled by pulsing $\Upsilon$) will map the
accumulated   phase onto population of singlet and triplet pairs.
To know $\phi$ is important as it determines the direction of the
anti-ferromagnetic order. Furthermore, by combining the blockade with
noise correlation measurements \cite{Altman} it is possible to
obtain further information about the magnetic phases. While the blockade probes  local correlation in the
DWs, noise measurements probe non-local spin-spin correlations  and thus can reveal long range order.

Before proceeding we note that similar  ideas  to that outlined
above can be used  for bosonic atoms if initially  no
$|t_{x,y}\rangle$ states are populated. The latter can be done by
detuning the $|t_{x,y}\rangle$ states by means of an external
magnetic field. In the bosonic case the doubly occupied $t_z$ states
will be the ones that  have the lowest energy. They will be
separated by an energy $\hbar \omega_0$ from the doubly occupied
singlets as the latter are the ones that have antisymmetric orbital
wave function in bosons. Consequently,  the role of $|s\rangle$ in
fermions will be replaced by $|t_z \rangle$ in bosons. The read-out
procedure would then be identical to that described above, while the
coherent dynamics will be given by the Hamiltonian Eq.(\ref{efJ}) apart from
the sign change $\zeta \rightarrow -\zeta$.

Up to now our analysis  has ignored tunneling between different DWs,
but in practice this tunneling can be controlled
 by tuning the lattice potential.  How will singlet and triplet pairs evolve due to this coupling?
We will now discuss the many-body dynamics that emerges when nearest
neighbor DW tunneling is allowed, i.e. $t_\sigma>0$.  When atoms can hop
between DWs, the behavior of the system will depend on the
dimensionality. For simplicity we will restrict our analysis  to a
1D array of $N$ double-wells, where  $t_\sigma$ corresponds to
hopping  energy of $\sigma$-type atoms between the right site of the $j^{th}-DW$
and the left site of the $(j+1)^{th}-DW$.

 In the regime $J_{\sigma}, t_{\sigma}, \Delta_\sigma\ll U$,
multiply occupied wells are energetically suppressed and  the
effective Hamiltonian is given by $\hat{H}^{eff}=
\hat{H}_{J}+\hat{H}_{t} $ . Here the first term corresponds to  the sum over $N$
independent $H_j^J$ Hamiltonians (see Eq.(\ref{efJ})),
$\hat{H}_{J}=\sum_{j=1}^N H_j^J$, each of which  acts on its
respective $j^{th}$-DW.  On the other hand $\hat{H}_{t}$  is
non-local as it couples different DWs and quartic as it
consists of  terms with four  singlet-triplet operators
\cite{future}.   The coupled DWs system  is  in general
 complex and  the quantum spin dynamics can be studied only numerically. However, there are specific
 parameter regimes where an exact solution can be found. For this
 discussion we will set  $\Delta_\sigma=0$. If  $t_{\uparrow}/t_{\downarrow} \to 0$, and at time $\tau=0$, no
$|t_x\rangle, |t_y\rangle $ triplet states are populated, their population will remain always
zero. Consequently, in this limit, the relevant Hilbert space
reduces to that of an effective  spin one-half system  with
$|s\rangle$ and $|t_z\rangle$ representing the effective $\pm 1/2$
states, which we  denote as $|\Uparrow\rangle$ and
$|\Downarrow\rangle$.  $\hat{H}_{t}$   couples such effective spin states. In the restricted Hilbert space
$\hat{H}^{eff}$ maps exactly to an
 Ising  chain in a magnetic field:
\begin{equation}
\addtolength{\belowdisplayskip}{-0.25
 cm}\addtolength{\abovedisplayskip}{-0.3 cm}
\hat{H}^{eff}= \mp\zeta \sum_{j}\hat{\sigma}_{ j}^{z} -\lambda_z
\sum_{j}\hat{\sigma}_{ j}^{x}\hat{\sigma}_{j+1}^{x}\label{Ising}
\end{equation} where $\hat{\sigma}^{\alpha}$ are the usual Pauli matrices which act of the effective  $|\Uparrow\rangle$ and
$|\Downarrow\rangle $ spins. In terms of singlet-triplet operators
they  are given by
 $\hat{\sigma}_{j}^{z}=(\hat{s}_{j}^\dagger\hat{s}_{j}-\hat{t}_{zj}^\dagger\hat{t}_{zj})$,
$\hat{\sigma}_{j}^{x}=\hat{s}_{j}^\dagger\hat{t}_{zj}+\hat{t}_{zj}^\dagger\hat{s}_{j}$
and
$\hat{\sigma}_{j}^{y}=(\hat{s}_{j}^\dagger\hat{t}_{zj}-\hat{t}_{zj}^\dagger\hat{s}_{j})/i$.
Here $\lambda_z=\frac{ t_{\downarrow}^2}{2 U}-\frac{
t_{\downarrow}^2 }{ U_{\downarrow\downarrow}}$  and the upper and
lower signs are for fermions and bosons respectively. For fermions  in the lowest vibrational level
the  onsite interaction energy between the same type of atoms
$U_{\uparrow\uparrow},U_{\downarrow\downarrow} \to \infty$ due to
the Pauli exclusion principle.

The 1D quantum Ising model  exhibits a second order quantum phase
transition at the critical value $|g|\equiv|\lambda_z/\zeta|=1$. For fermions (upper sign) when
$g\ll 1$ the ground state corresponds to  all effective spins
pointing up, i.e $|G\rangle=|\Uparrow \dots \Uparrow \rangle=\Pi
_j|s\rangle_j$. On the other hand when $g \gg 1$, there are two
degenerate ground states which are, in the effective spin basis,
macroscopic superpositions of oppositely polarized states along $x$.
In terms of the original fermionic spin states this superposition
correspond to the states
 $|AF^{\pm}\rangle= \frac{1}{\sqrt{2}}( |\uparrow\downarrow\dots \uparrow\downarrow\rangle\pm
|\downarrow\uparrow\dots \downarrow\uparrow \rangle )$. Therefore,
by adiabatic passage one could start with $|G\rangle$ and convert it
into  AF state(s). Due  to vanishing energy gap at the quantum critical
point $g=1$, adiabaticity is difficult to maintain as  $N \to
\infty$ \cite{Zurek,Polkovnikov,Cherng,Dziarmaga}. In that respect,
our projection scheme is useful to test adiabatic following. It can
be done either by measuring the number of
$|\uparrow\downarrow\rangle$ pairs in the final state or by adiabatically
ramping down $g$ back   to zero
 and  measuring  the number of singlets/triplet pairs. The remaining  number of triplets will
determine  the number of excitations created in  the  process.

%After the AF state is created, one can turn off the tunneling between ow well an  $|AF\rangle$ state is prepared. For example, ones it has been created,  one can slowly decrease  $g$  back   to zero, and  measure  the number of singlets/triplet pairs. The remaining number of triplets will probe  the number of excitations created during  the  process.

We now turn to non-adiabatic dynamics. We will discuss the situation
where initially  the system is prepared  in a product of singlet
states ($\lambda_z=0$ ground state ) and then one  lets  it evolve
for $\tau>0$  with a fixed $|\lambda_z|>0$. Generically the coupling
between DWs results in oscillations between singlet and triplet
pairs with additional decay on a slower time scale. We present two
important special cases involving such dynamics:

i) {\it Singlet-triplet cluster state generation:} If the value of
$\lambda_z$ is set to be $|\lambda_z|\gg \zeta$, then the Hamiltonian
reduces to a pure Ising Hamiltonian and thus at particular times,
$\tau_c$, given by $\lambda_z \tau_c/\hbar= \pi/4 $ mod $\pi/2$
the evolving state becomes a $d=1$ cluster state $|\mathcal
C\rangle$ in the effective spin basis \cite{Brigel}. Up to single spin rotations
$|\mathcal{C}\rangle=\frac{1}{2^{N/2}}
\bigotimes_{j=1}^{N}(|\Uparrow\rangle_j
\hat{\sigma}_{j+1}^{z}+|\Downarrow \rangle_j)$.
Cluster states are of interest for the realization of one-way quantum computation proposals where starting  from the state
$|\mathcal
C\rangle$
 computation can be done via measurements only. Preparation of cluster states encoded in the logical
$\Uparrow,\Downarrow$ qubits  may have  significant practical advantages since the $\Uparrow,\Downarrow$ states  have  zero net
spin  along the quantization axis  and hence are not affected by  global magnetic field
fluctuations. Additionally, the use  of such  singlet-triplet states for encoding  might
allow for the generation of  decoherence free subspaces
insensitive to  collective and local
 errors \cite{Weinstein} and for  alternative schemes for measured-based
quantum computation \cite{Gross}.

ii) {\it Non-equilibrium generation  and probing of AF
correlations:} The second situation is when the value of $\lambda_z$
is set to the critical value, $|\lambda_z|=\zeta$ (or $g=1$). We
will first focus on the fermionic system $\lambda_z>0$. To discuss
it,   we remind that the dynamics driven by $\hat{H}^{eff}$ is
exactly solvable as  $\hat{H}^{eff}$ can be mapped via  the Jordan Wigner transformation   into a quadratic
Hamiltonian of fermionic operators which  can be diagonalized by a
canonical transformation \cite{JWT, Dziarmaga}. Using such
transformation  it is possible to show that  at specific times, the shortest of them we denote by $\tau_{m}\approx \hbar
\frac{N+1}{4 \zeta}$, long range AF correlations build up and for
small atom number the state approaches $|AF^+\rangle$. To quantify
the resulting state  in Fig.~2(inset) we plot the fidelity, defined
as $\mathcal{F}_1(\tau_m)=|\langle
AF^+|\psi(\tau_m)\rangle{}_{g=1}|^2$, as a function of $N$. The
figure shows that while an almost perfect $|AF^+\rangle$ is
dynamically generated  for small $N$, its fidelity exponentially
degrades with increasing  atom number.

However, the fidelity is a very strict probe, as it drops to zero
when a single spin is flipped. As $N$ increases the system ends at
$\tau_m$ in a quantum superposition of states like $ |\dots
\Rightarrow\Leftarrow\Leftarrow\Leftarrow\Leftarrow\Leftarrow\Rightarrow\Rightarrow\Rightarrow\Rightarrow\Rightarrow\Rightarrow
\Leftarrow~\dots~\rangle$ with finite domains of "effective spins"
pointing along $\pm x$, separated by kinks where the polarization of
the spins change its orientation (we used the convention
$|\uparrow\downarrow\rangle\equiv|\Rightarrow\rangle$).
Consequently, one gets more realistic information about the AF order
of the state, by measuring the average size of the domains or the
average density of kinks, the latter defined as
$\nu\equiv\frac{1}{2N}\sum_j(1-\langle \psi
(\tau)|\hat{\sigma}^x_j\hat{ \sigma}^x_{j+1}| \psi (\tau)\rangle)$.

Our read-out technique can be used to detect the kink-density as for
an arbitrary fixed $g$  energy conservation imposes a relation
between  $\nu$ and the triplet-z density, $N_t$:\begin{equation}
\addtolength{\belowdisplayskip}{-0.25
 cm}\addtolength{\abovedisplayskip}{-0.3 cm}
\nu(\tau,g)=\frac{1}{2}-\frac{N_t(\tau,g)}{g}.\label{deff}
\end{equation} A simple analytical
expression for $N_t(\tau,g)$ can be obtained by using the Jordan Wigner transformation \cite{JWT}: $N_t(\tau,g)=\frac{1}{N}\lambda_z^2
\sum_{k=0}^{N-1}\frac{\sin^2(2\pi k/N) \sin^2(2 \omega_k\tau)}
{\hbar^2 \omega_k^2}$ where
$\hbar\omega_k=\zeta\sqrt{g^2+1+2g\cos(2\pi k/N)}$ are
 quasi-particle frequencies of $\hat{H}^{eff}$. The fact that
  it remains always below $0.2$ (see Fig. 2) confirms the
 idea that regardless of the reduced fidelity at large $N$, the
state does retain AF correlations. We point out that $|AF^+\rangle$ states
are only generated at  $g=1$, a feature that illustrates the special
character of the critical dynamics.

\begin{figure}[htbp]
\centering \includegraphics[width=3.3 in,height=1.9 in]{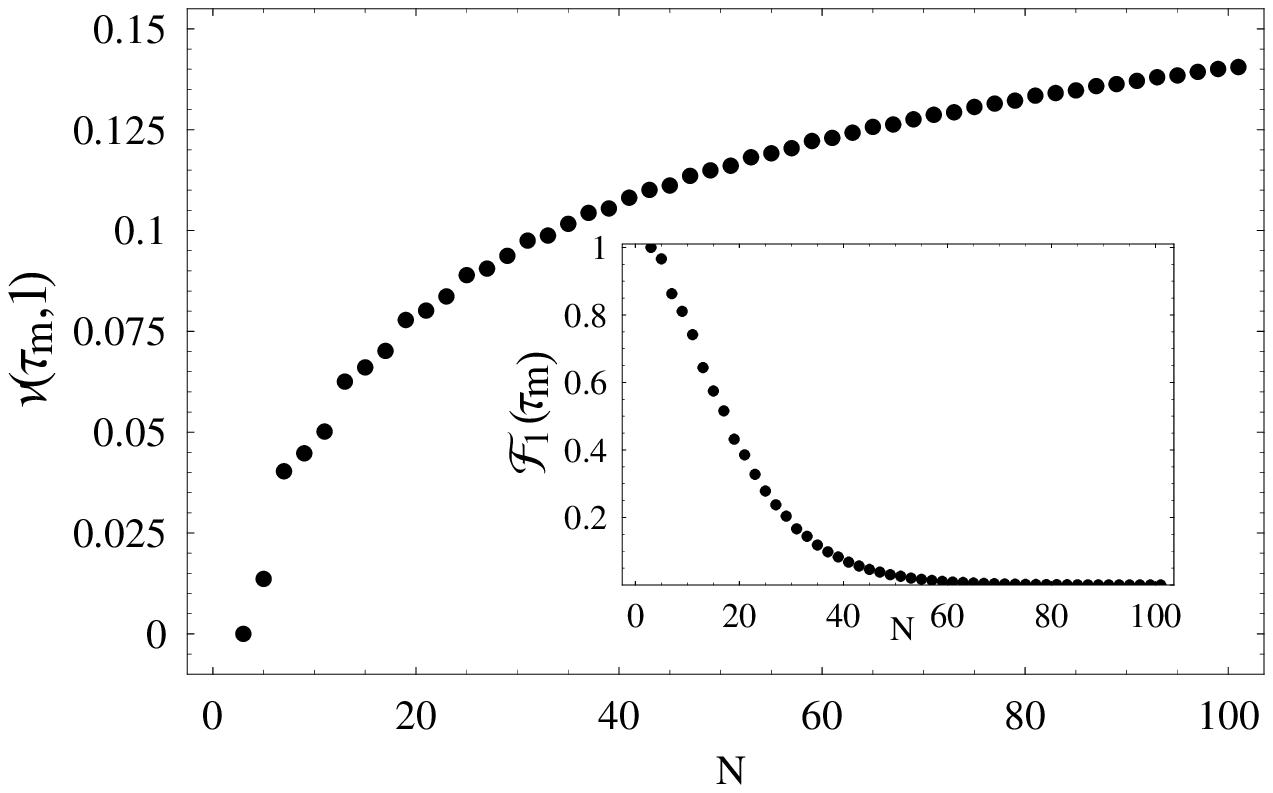}
\caption{Using the the Jordan-Wigner transformation \cite{JWT} we calculated the density of kinks {\it vs} $N$ at $\tau=\tau_m$
and the fidelity
$|\langle\psi(\tau_m)|AF^+\rangle|^2$ {\it vs} $N$ (inset).
Our projection
technique can  be used to measure $\nu(\tau)$ as it is directly
related to the triplet density, $N_t(\tau)$ (see Eq.(7)). }
\end{figure}

Let us  now discuss the bosonic case. If $\lambda_z>0$,
the  fermionic results apply  for bosons by simply
interchanging the role of $|s\rangle\leftrightarrow|t_z\rangle$. On the other hand if $\lambda_z<0$, not only one has to interchange
 $|s\rangle\leftrightarrow|t_z\rangle$ but additionally,
 the adiabatic and non-equilibrium dynamics will generate, instead of  $|AF^{\pm}\rangle$ states,
$ \frac{1}{\sqrt{2}}( |\Rightarrow\Leftarrow\dots \Rightarrow\Leftarrow\rangle \pm
|\Leftarrow\Rightarrow\dots \Leftarrow\Rightarrow \rangle )$ i.e
macroscopic superpositions of
   AF states along
the $x$-direction  in the effective spin basis. With these modifications, the  results  derived for fermions  hold for bosons\footnote{ In this case a
different sign   in the definition of kink density
$\hat{\sigma}^x_j\hat{ \sigma}^x_{j+1} \to -\hat{\sigma}^x_j\hat{
\sigma}^x_{j+1}$ is required.}.

Before concluding we   briefly mention that spin dependent superlattices of the form
\begin{equation}
\addtolength{\belowdisplayskip}{-0.25
 cm}\addtolength{\abovedisplayskip}{-0.3 cm}
V= \sum_{j=1,2}(A_j+B_j\sigma_z ) \cos^{2}[k z/j+\theta_j]
\end{equation} can be experimentally realized  by superimposing two independent
lattices, generated by elliptically polarized light, one with twice the periodicity of the
other  \cite{Peil,Sebby,Porto}. Complete control over the DW parameters
 is achieved by controlling the
 phases (which determine $\Delta$), intensities (which determine $U$,$J$ and $t$) and  polarization of the
 laser beams (which allow for spin dependent control). For example  lattice configurations with $t_\uparrow\ll t_\downarrow$
can be achieved  by setting the laser parameters such  that $B_1=0$  and  $A_2=B_2 \gg 1$.

In summary we have described a technique to prepare, detect and
manipulate spin  configurations in ultra-cold  atomic systems loaded
in spin dependent  period-two superlattices. By studying the
many-body dynamics that arises when tunneling between DWs is
allowed, we discussed how to dynamically generate singlet-triplet
cluster states and  AF cat states, which are of interest for
quantum information science,  and how to probe AF correlations in
far from equilibrium dynamics. Even though in this Letter we
restrict our analysis to 1D systems the ideas developed here can
be extended to higher dimensions and more general kinds of
interactions.

We acknowledge useful discussions with G. Morigi.
This work was supported by ITAMP, NSF (Career Program), Harvard-MIT CUA, AFOSR, Swiss
NF, the Sloan Foundation, and the David and Lucille Packard
Foundation.

\addtolength{\parskip}{-0.6 cm}

\begin{spacing}{0.98}
\bibliographystyle{plain}

\end{spacing}

\end{document}